\documentclass[epj,nopacs]{svjour}
\usepackage{amssymb}
\usepackage{epsfig}
\newcommand{\directprod}{\mathop{\otimes}}

\begin{document}

\title{Yukawa Coupling Structure in Intersecting D-brane Models}

\author{
Noriaki Kitazawa\inst{1}
\and
Tatsuo Kobayashi\inst{2}
\and
Nobuhito Maru\inst{3}
\and
Nobuchika Okada\inst{4}
}

\institute{
Department of Physics, Tokyo Metropolitan University,
Hachioji, Tokyo 192-0397, Japan\\
\email{kitazawa@phys.metro-u.ac.jp}
\and
Department of Physics, Kyoto University, Kyoto 606-8502, Japan\\
\email{kobayash@gauge.scphys.kyoto-u.ac.jp}
\and
Theoretical Physics Laboratory, RIKEN,
2-1 Hirosawa, Wako, Saitama 351-0198, Japan\\
\email{maru@postman.riken.go.jp}
\and
Theory Group, KEK,
1-1 Oho, Tsukuba, Ibaraki 305-0801, Japan\\
\email{okadan@post.kek.jp}
}

\date{}

\abstract{
The structure of Yukawa coupling matrices is investigated
 in type IIA $T^6/({\rm Z}_2 \times {\rm Z}_2)$
 orientifold models with intersecting D-branes.
 Yukawa coupling matrices are difficult to be realistic
 in the conventional models in which the generation structure
 emerges by the multiple intersection of D-branes
 in the factorized $T^6 = T^2 \times T^2 \times T^2$.
We study the new type of flavor structure, where 
Yukawa couplings are dynamically generated, 
and show this type of models lead to nontrivial structures of 
Yukawa coupling matrices, which can be realistic. 
\PACS{
      {PACS-key}{discribing text of that key}   \and
      {PACS-key}{discribing text of that key}}
}

\maketitle

\section{Introduction}
\label{sec:intro}

Understanding the masses and flavor mixing of quarks and leptons
 is one of the most important issues in particle physics.
New physics
 which derive Yukawa coupling matrices in the standard model
 should certainly exist.
Yukawa couplings, in a sense, seem naturally of ${\cal O}(1)$.
Thus, how to derive suppressed Yukawa couplings is 
a key to understand the hierarchy of fermion masses and 
mixing angles.

String theory,
 which is a candidate of the consistent theory of quantum gravity,
 gives an attractive framework to generate Yukawa coupling matrices.
Indeed, in several types of string models coupling selection rules 
have been studied and Yukawa couplings have been calculated.
Recent development of the models with intersecting D-branes,
 intersecting D-brane models
 (see Refs.\cite{BGKL,AFIRU,Blumenhagen:2000ea,CSU} for essential idea),
 opens a new possibility towards realistic models.
Among them,
 models with low-energy supersymmetry
 \cite{CSU,Forste:2000hx,BGO,Cvetic:2002pj,Honecker,Larosa-Pradisi,Cvetic-Papadimitriou,kitazawa1,Cvetic:2004ui,Honecker-Ott,CLS,kitazawa2}
 are interesting,
 because those are constructed as stable solutions
 of the string theory.
Within the framework of intersecting D-brane models, 
open string modes corresponding to matter fields 
and Higgs fields are localized at intersecting points 
between D-branes \cite{Berkooz:1996km}.
The flavor number is obtained as intersecting numbers of 
D-branes in conventional models.

Intersecting D-brane models have several interesting 
aspects from phenomenological viewpoints.
In particular, their Yukawa couplings have a 
phenomenologically important feature. 
Calculation of Yukawa couplings in 
intersecting D-brane models
 \cite{CIM,Cvetic:2003ch,Abel:2003vv,LMRS,Cremades:2004wa}
is similar to Yukawa calculation of 
twisted strings in 
heterotic orbifold models
 \cite{Dixon:1986qv,Burwick:1990tu,Kobayashi:2003vi}
 \footnote{
Yukawa couplings in heterotic models are calculable, 
because string is solvable on orbifolds.}.
These Yukawa couplings in 
both intersecting D-brane models and heterotic orbifold models 
depend on localized points of 
matter modes and moduli corresponding 
to the volume of compact space. 
Thus, Yukawa couplings are, in general, non-universal, and 
can lead to suppressed values.
Indeed, heterotic orbifold models have possibilities for 
realizing realistic Yukawa matrices \cite{Ko:2004ic}.
Also intersecting D-brane models have the potential 
to lead to realistic Yukawa matrices, but 
most of intersecting D-brane models, 
which have been constructed so far, seem to lead to 
the factorizable form of Yukawa matrices, $y_{ij} = a_i b_j$, 
that is, rank-one matrices.
With this form of Yukawa matrices, one can derive 
only a non-vanishing mass for the third family, but 
vanishing values for other lighter masses and mixing angles \cite{Abel:2003yh}.
Hence, it is quite important to study a new type of 
flavor structure which can lead to non-vanishing mixing 
angles and light fermion masses within the framework of 
intersecting D-brane models.

In Ref.\cite{kitazawa2}
 one of the authors has proposed an intersecting D-brane model with 
a new type of flavor structure, where 
quarks, leptons and Higgs doublets appear as composite fields.
In this model, Yukawa couplings are dynamically generated.
Since 
 the origin of flavors is different from the one
 in the conventional models,
 the structure of Yukawa coupling matrices can be different
 from the one in conventional models.
In this paper
 we investigate 
 the structure of Yukawa coupling matrices of the model,
 and show that the nontrivial form of Yukawa matrices is 
 obtained in this framework of Yukawa coupling generations, 
 e.g. non-vanishing mixing angles.
Thus, our scenario is interesting to realize 
realistic values of fermion masses and 
mixing angles.

This paper is organized as follows.
In section \ref{sec:review}, we give a brief review on intersecting 
D-brane models.
Then, 
 we discuss the structure of Yukawa coupling matrices
 in conventional models.
We point out
 that it is difficult to have realistic structure in general.
In section \ref{sec:model} we briefly introduce
 a model of dynamical generation of Yukawa couplings
 which is proposed in Ref.\cite{kitazawa2}.
In section \ref{sec:structure}
 a detailed analysis of the structure of Yukawa coupling matrices
 is given.
We show that
 our framework of the Yukawa coupling generation derives 
 a nontrivial structure of Yukawa matrices, which can be realistic.
Section \ref{sec:conclusions} is devoted to 
conclusions and discussions.

\section{Intersecting D-brane models and Yukawa couplings}
\label{sec:review}

\subsection{Intersecting D-brane models}

Here we give a brief review on intersecting D-brane models.
Consider the type IIA superstring theory compactified on
 ${\bf T}^6/({\bf Z}_2 \times {\bf Z}_2)$ orientifold, where
 ${\bf T^6}={\bf T^2} \times {\bf T^2} \times {\bf T^2}$.
The type IIA theory is invariant
 under the ${\bf Z_2} \times {\bf Z_2}$ transformation
\begin{eqnarray}
 \theta: && \qquad X_{\pm}^k \rightarrow e^{\pm i 2\pi v_k} X_{\pm}^k,
\\
 \omega: && \qquad X_{\pm}^k \rightarrow e^{\pm i 2\pi w_k} X_{\pm}^k,
\end{eqnarray}
 where $v=(0,0,1/2,-1/2,0)$ and $w=(0,0,0,1/2,-1/2)$ and
\begin{equation}
 X_{\pm}^k = 
  \left\{
   \begin{array}{ll}
    {1 \over \sqrt{2}} \left( \pm X^{2k} + X^{2k+1} \right),
     & \quad \mbox{for $k=0$}, \\
    {1 \over \sqrt{2}} \left( X^{2k} \pm i X^{2k+1} \right),
     & \quad \mbox{for $k=1,2,3,4$}
   \end{array}
  \right.
\end{equation}
 with space-time coordinates $X^\mu$, $\mu=0,1,\cdots,9$.
The type IIA theory
 is also invariant under the $\Omega R$ transformation,
 where $\Omega$ is the world-sheet parity transformation and
\begin{equation}
 R: \qquad\quad
  \left\{
   \begin{array}{ll}
    X^i \rightarrow X^i,
     & \quad \mbox{for $i=0,1,2,3,4,6,8$}, \\
    X^j \rightarrow -X^j,
     & \quad \mbox{for $j=5,7,9$}.
   \end{array}
  \right.
\end{equation}
We mod out the theory
 by the action of $\theta$, $\omega$, $\Omega R$
 and their independent combinations.

A D6${}_a$-brane stretching over our three-dimensional space
 and winding in compact
 ${\bf T^2} \times {\bf T^2} \times {\bf T^2}$ space
 is specified by the winding numbers in each torus:
\begin{equation}
 [(n_a^1, m_a^1), (n_a^2, m_a^2), (n_a^3, m_a^3)].
\end{equation}
A D6${}_a$-brane is always accompanied by its orientifold image
 D6${}_{a'}$ whose winding numbers are
\begin{equation}
 [(n_a^1, -m_a^1), (n_a^2, -m_a^2), (n_a^3, -m_a^3)].
\end{equation}
The number of intersection
 between D6${}_a$-brane and D6${}_b$-brane is given by
\begin{equation}
 I_{ab} = \prod_{i=1}^3 \left( n_a^i m_b^i - m_a^i n_b^i \right).
\end{equation}
The intersecting angles, $\theta_a^i$ with $i=1,2,3$,
 between D6${}_a$-brane and $X^4$, $X^6$ and $X^8$ axes
 in each torus are given by
\begin{equation}
 \theta_a^i = \tan^{-1} \left( \chi_i {{m_a^i} \over {n_a^i}} \right),
\end{equation}
 where $\chi_i$ are the ratios of two radii of each torus:
 $\chi_i \equiv R^{(i)}_2 / R^{(i)}_1$.
The system has supersymmetry,
 if $\theta_a^1+\theta_a^2+\theta_a^3=0$ is satisfied for all $a$.
The configuration of intersecting D6-branes should satisfy
 the following Ramond-Ramond tadpole cancellation conditions, 
\begin{eqnarray}
 \sum_a N_a n_a^1 n_a^2 n_a^3 &=& 16,
\\
 \sum_a N_a n_a^1 m_a^2 m_a^3 &=& -16,
\\
 \sum_a N_a m_a^1 n_a^2 m_a^3 &=& -16,
\\
 \sum_a N_a m_a^1 m_a^2 n_a^3 &=& -16,
\end{eqnarray}
 where $N_a$ is the multiplicity of D6${}_a$-brane,
 and we are assuming three rectangular (untilted) tori.
The Neveu-Schwarz-Neveu-Schwarz tadpoles are automatically cancelled,
 when Ramond-Ramond tadpole cancellation conditions and
 the supersymmetry conditions are satisfied.

There are four sectors of open string corresponding to
 on which D6-branes two ends of open string are fixed:
 $aa$, $ab+ba$, $ab'+b'a$, $aa'+a'a$ sectors.
Each sector gives matter fields
 in four-dimensional low-energy effective theory.
The general massless field contents
 are given in Table \ref{general-spectrum}.
\begin{table}
 \begin{tabular}{|c|l|}
  \hline
  sector & field \\
  \hline\hline
  $aa$   & U$(N_a/2)$ or USp$(N_a)$ gauge multiplet. \\
         & 3 U$(N_a/2)$ adjoint or 3 USp$(N_a)$ \\
         & anti-symmetric tensor chiral multiplets.\\
  \hline
  $ab+ba$ & $I_{ab}$ $( \Box_a, {\bar \Box}_b )$ chiral multiplets. \\
  \hline
  $ab'+b'a$ & $I_{ab'}$ $( \Box_a, \Box_b )$ chiral multiplets. \\
  \hline
  $aa'+a'a$ & ${1 \over 2}
                ( I_{aa'} - {4 \over {2^k}} I_{aO6} )$
                symmetric tensor \\
            &   chiral multiplets. \\
            & ${1 \over 2}
                ( I_{aa'} + {4 \over {2^k}} I_{aO6} )$
                anti-symmetric tensor \\
            &   chiral multiplets. \\
  \hline
 \end{tabular}
\caption{
General massless field contents on intersecting D6-branes.
In $aa$ sector,
 the gauge symmetry is USp$(N_a)$ or U$(N_a/2)$ corresponding to
 whether D6${}_a$-brane is parallel or not
 to some O6-plane, respectively.
In $aa'+a'a$ sector, $k$ is the number of tilted torus,
 and $I_{aO6}$ is the sum of the intersection numbers
 between D6${}_a$-brane and all O6-planes.
}
\label{general-spectrum}
\end{table} 
A common problem of the model building in this framework
 is the appearance of the massless adjoint fields in $aa$ sector,
 since there are no massless matter fields
 in the adjoint representation
 under the standard model gauge group in Nature.
These fields are expected to be massive
 in the case of a curved compact space,
 because these are the moduli fields of D6-brane configurations.

\subsection{The Yukawa Coupling Structure in Conventional Models}

The magnitude of Yukawa couplings among localized 
open string modes can be calculated by conformal 
field theory technique like heterotic orbifold models.
The important part of 3-point interactions $y$ is 
evaluated by the classical part 
as \cite{CIM,Cvetic:2003ch,Abel:2003vv}
\begin{equation}
y \sim e^{-S_{cl}} \sim \prod_i e^{-\Sigma_i/2\pi \alpha'}.
\label{Yukawa-cl}
\end{equation}
Here, $S_{cl}$ denotes the action of 
classical string solution $X_{cl}$, which have the asymptotic 
behavior corresponding to local open string modes 
near intersecting points, and 
$S_{cl}$ is obtained as a product of the triangle areas $\Sigma_i$ 
on the $i$-th $T^2$ , which 
string sweeps to couple.
These Yukawa couplings can lead to suppressed 
values when intersecting points are far away each other.
Thus, intersecting D-brane models have the potential 
to realize the hierarchy of fermion masses and mixing angles.

Here, however, 
 we discuss the difficulty to generate
 a realistic structure of Yukawa coupling matrices
 in the models from type IIA $T^6/({\rm Z}_2 \times {\rm Z}_2)$
 orientifolds with intersecting D-branes.
The six-dimensional torus is assumed to be factorizable:
 $T^6 = T^2 \times T^2 \times T^2$.

As an illustrating example, let us consider 
a toy model with three D6-branes,
 D6${}_3$-brane for SU$(3)_c$ gauge symmetry,
 D6${}_2$-brane for SU$(2)_L$ gauge symmetry and
 D6${}_1$-brane for one component of U$(1)_Y$ gauge symmetry,
 and a very simple configuration of intersecting numbers
\begin{equation}
 I_{ab} =
 \left(
  \begin{array}{ccc}
    0 &  1 & 3 \\
   -1 &  0 & 3 \\
   -3 & -3 & 0
  \end{array}
 \right)_{ab},
\end{equation}
 where $a,b = 1,2,3$.
This system gives three generations of quarks,
 three left-handed quark doublets
 and three right-handed quark singlets,
 and one Higgs doublet with Yukawa couplings.
The total intersecting number
 $I_{ab}$ of D6${}_a$-brane and D6${}_b$-brane
 is given by the multiplications of three intersecting numbers
 of D6${}_a$-brane and D6${}_b$-brane in each of three tori.
The structure of Yukawa coupling
 is not fully determined by the intersecting number,
 but it depends on the configuration of the intersection
 in each of three tori.

The intersecting number $I_{12}=1$ means
 one Higgs doublet field localizing a point in each of three tori.
The intersecting number $I_{13}=3$ means
 three intersections in one of three tori
 and one intersection in other two tori.
The same is true for $I_{23}=3$.
Therefore,
 we have two different configurations of intersection:
 (1) D6${}_1$-brane and D6${}_3$-brane intersect three times
     in a torus in which
     D6${}_2$-brane and D6${}_3$-brane also intersect three times,
 or
 (2) D6${}_1$-brane and D6${}_3$-brane intersect three times
     in a torus in which
     D6${}_2$-brane and D6${}_3$-brane intersect once.

In both cases, Yukawa couplings are evaluated by Eq.(\ref{Yukawa-cl}).
The quantum part also contributes to Yukawa couplings, but gives 
${\cal O}(1)$ of common factor.
In case (1)
 the Yukawa matrix can become hierarchical but always diagonal.
In this case
 D6${}_3$-brane should wind three times in a torus
 and intersect with each D6${}_1$-brane and D6${}_2$-brane
 once in each winding. 
This gives
 definite three pairs of right-handed and left-handed quarks
 and defines three generations without mixing.
The values of Yukawa couplings
 are determined by the areas of triangles
 which are determined by the place of each pair of quarks
 and Higgs doublet.
In this case we can not have flavor mixing.
This structure of Yukawa coupling matrix
 seems unable to be modified by the quantum corrections 
 in supersymmetric models.

In case (2)
 we always have factorized structure of Yukawa coupling matrix:
\begin{equation}
 y_{ij} \propto a_i b_j,
\label{yukawa-fact}
\end{equation}
 where $i,j = 1,2,3$.
In a torus where
 D6${}_1$-brane intersects three times with D6${}_3$-brane
 three left-handed quarks localize at one point
 and Higgs doublet also localizes at one point.
We have three triangle areas
 which determine the value of $a_i$ or $b_i$.
The same is true for the torus where
 D6${}_2$-brane intersects three times with D6${}_3$-brane.
Therefore,
 the Yukawa coupling matrix is the multiplication
 of the contribution from each three torus.
Furthermore, the K\"ahler metric depends on twisted angles of 
open string \cite{LMRS}.
The twisted angle of open strings corresponding to 
three left-handed quarks is common, and we also have the same twisted angle 
for three right-handed quarks.
Thus, the K\"ahler metric is relevant to only the overall magnitude 
of Yukawa matrices, but irrelevant to ratios of entries in 
Yukawa matrices.
In this case
 we can not have realistic mass spectrum,
 since the rank of the Yukawa coupling matrix is one.
This structure of Yukawa coupling matrix
 also seems unable to be modified by the quantum corrections 
 in supersymmetric models. 

We have discussed the problem in the simplified model, but 
most of models seem to have the almost same problem.
In particular, it seems difficult to derive non-vanishing 
mixing angles as well as lighter fermion masses.
There are several possibilities to overcome this difficulty.

First,
 we can try to construct models with some further complications
 by arranging the configuration of D6-branes.
For example, 
 introduction of many Higgs doublets may solve the problem
 \cite{Chamoun:2003pf},
 though it is not experimentally favored.
To have three generations
 utilizing orientifold image D6-branes
 may give a way to solve the problem,
 though we will have relatively many unwanted exotic fields.

The second way is
 to change the structure of the compactified space.
Leaving from the factorizable $T^6 = T^2 \times T^2 \times T^2$
 to $T^6 = T^4 \times T^2$, for example,
 may make the situation completely change.
But the physics becomes less intuitive
 and the analysis becomes much more complicated.

The third way is to change the origin of the generation.
This is the way we are going to take in this paper.
In the next section,
 we introduce a model in which
 the generation is not originated from
 the multiple intersection of D6-branes.

\section{A Model of Dynamically Generated Yukawa Couplings}
\label{sec:model}

In this section
 we give a brief description of the model
 which has been introduced in Ref.\cite{kitazawa2}.

The D6-brane configuration of the model
 is given in Table \ref{config}.
\begin{table}
 \begin{tabular}{|c|c|c|}
  \hline
  D6-brane & winding number & multiplicity     \\
  \hline\hline
  D6${}_1$   & $ \quad [(1,-1), (1,1), (1,0)] \quad $ & $4$ \\
  \hline
  D6${}_2$   & $ \quad [(1,1), (1,0), (1,-1)] \quad $ & $6+2$  \\
  \hline
  D6${}_3$   & $ \quad [(1,0), (1,-1), (1,1)] \quad $  & $2+2$  \\
  \hline
  D6${}_4$   & $ \quad [(1,0), (0,1), (0,-1)] \quad $ & $12$  \\
  \hline
  D6${}_5$   & $ \quad [(0,1), (1,0), (0,-1)] \quad $ & $8$ \\
  \hline
  D6${}_6$   & $ \quad [(0,1), (0,-1), (1,0)] \quad $ & $12$ \\
  \hline
 \end{tabular}
\caption{
Configuration of intersecting D6-branes.
All three tori are considered to be rectangular (untilted).
Three D6-branes, D6${}_4$, D6${}_5$ and D6${}_6$,
 are on top of some O6-planes.
}
\label{config}
\end{table}
Both tadpole cancellation conditions and supersymmetry conditions
 are satisfied in this configuration under the conditions of
 $\chi_1 = \chi_2 = \chi_3 \equiv \chi$.
The D6${}_2$-brane system consists of
 two parallel D6-branes with multiplicities six and two
 which are separated in the second torus
 in a consistent way with the orientifold projections.
The D6${}_3$-brane system consists of
 two parallel D6-branes with multiplicity two
 which are separated in the first torus
 in a consistent way to the orientifold projections.
D6${}_1$, D6${}_2$ and D6${}_3$ branes give gauge symmetries
 of U$(2)_L=$SU$(2)_L \times$U$(1)_L$,
 U$(3)_c \times$U$(1) =$SU$(3)_c \times$U$(1)_c \times$U$(1)$
 and U$(1)_1 \times$U$(1)_2$,
 respectively.
The hypercharge is defined as
\begin{equation}
 {Y \over 2} = {1 \over 2} \left( {{Q_c} \over 3} - Q \right)
             + {1 \over 2} \left( Q_1 - Q_2 \right),
\end{equation}
 where $Q_c$, $Q$, $Q_1$ and $Q_2$ are charges of
 U$(1)_c$, U$(1)$, U$(1)_1$ and U$(1)_2$, respectively.
The additional non-anomalous U(1) charge, $Q_R$, is defined as
\begin{equation}
 Q_R = Q_1 - Q_2.
\end{equation}
The remaining three U$(1)$ gauge symmetries
 which are generated by
 $Q_L$ (namely U$(1)_L$), $Q_c+Q$ and $Q_1+Q_2$ are anomalous,
 and their gauge bosons have masses of the order of the string scale.
These three anomalous U$(1)$ gauge symmetries
 are independent from the two non-anomalous U$(1)$ gauge symmetries:
 ${\rm tr}((Y/2) Q_L) = 0$, for example.

\begin{figure}
\begin{center}
\epsfig{file=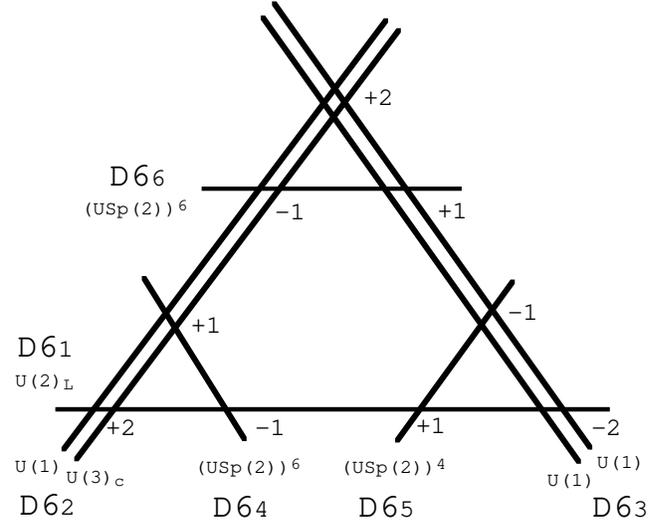,height=70mm}
\end{center}
\caption{
Schematic picture of the configuration of intersecting D6-branes.
This picture describes
 only the configuration of the intersection of 
D6-branes each other,
 and the relative place of each D6-brane has no meaning.
The number at the intersection point
 between D6${}_a$ and D6${}_b$ branes
 denotes intersection number $I_{ab}$ with $a<b$.
}
\label{intersec}
\end{figure}

A schematic picture
 of the configuration of intersecting D6-branes of this model
 is given in Fig.\ref{intersec}.
There are no $ab'+b'a$, $aa'+a'a$ sectors of open string
 in this configuration.
The massless particle contents are given in Table \ref{contents_1}.
\begin{table}
 \begin{tabular}{|c|c|c|}
  \hline
  sector             & $\mbox{SU}(3)_c \times \mbox{SU}(2)_L
                       \times \mbox{USp}(8)$ 
                     & field  \\
                     & $\times \mbox{USp}(12)_{D6_4}
                       \times \mbox{USp}(12)_{D6_6}$
                     & \\
                     & ($Y/2, Q_R$)($Q_L, Q_c+Q, Q_1+Q_2$)
                     &        \\
  \hline\hline
  $D6_1 \cdot D6_2$  & $(3^*, 2, 1, 1, 1)_{(-1/6,0)(+1,-1,0)}
                         \times 2$
                     & ${\bar q}_i$ \\
                     & $(1, 2, 1, 1, 1)_{(+1/2,0)(+1,-1,0)}
                         \times 2$
                     & ${\bar l}_i$  \\
  \hline
  $D6_1 \cdot D6_4$  & $(1, 2, 1, 12, 1)_{(0,0)(-1,0,0)}$
                     & $D$ \\
  \hline
  $D6_2 \cdot D6_4$  & $(3, 1, 1, 12, 1)_{(+1/6,0)(0,+1,0)}$
                     & $C$  \\
                     & $(1, 1, 1, 12, 1)_{(-1/2,0)(0,+1,0)}$
                     & $N$  \\
  \hline
  $D6_1 \cdot D6_3$  & $(1, 2, 1, 1, 1)_{(+1/2,+1)(-1,0,+1)}
                        \times 2$
                     & $H^{(1)}_i$ \\
                     & $(1, 2, 1, 1, 1)_{(-1/2,-1)(-1,0,+1)}
                        \times 2$
                     & ${\bar H}^{(2)}_i$ \\
  \hline
  $D6_1 \cdot D6_5$  & $(1, 2, 8, 1, 1)_{(0,0)(+1,0,0)}$
                     & $T$ \\
  \hline
  $D6_3 \cdot D6_5$  & $(1, 1, 8, 1, 1)_{(+1/2,+1)(0,0,-1)}$
                     & $T^{(+)}$ \\
                     & $(1, 1, 8, 1, 1)_{(-1/2,-1)(0,0,-1)}$
                     & $T^{(-)}$ \\
  \hline
  $D6_2 \cdot D6_3$  & $(3, 1, 1, 1, 1)_{(-1/3,-1)(0,+1,-1)}
                        \times 2$
                     & ${\bar d}_i$ \\
                     & $(3, 1, 1, 1, 1)_{(+2/3,+1)(0,+1,-1)}
                        \times 2$
                     & ${\bar u}_i$ \\
                     & $(1, 1, 1, 1, 1)_{(-1,-1)(0,+1,-1)}
                        \times 2$
                     & ${\bar e}_i$ \\
                     & $(1, 1, 1, 1, 1)_{(0,+1)(0,+1,-1)}
                        \times 2$
                     & ${\bar \nu}_i$ \\
  \hline
  $D6_2 \cdot D6_6$  & $(3^*, 1, 1, 1, 12)_{(-1/6,0)(0,-1,0)}$
                     & ${\bar C}$ \\
                     & $(1, 1, 1, 1, 12)_{(+1/2,0)(0,-1,0)}$
                     & ${\bar N}$ \\
  \hline
  $D6_3 \cdot D6_6$  & $(1, 1, 1, 1, 12)_{(+1/2,+1)(0,0,+1)}$
                     & ${\bar D}^{(+)}$ \\
                     & $(1, 1, 1, 1, 12)_{(-1/2,-1)(0,0,+1)}$
                     & ${\bar D}^{(-)}$ \\
  \hline
 \end{tabular}
\caption{
Low-energy particle contents before ``hypercolor'' confinement.
The fields from $aa$ sectors are neglected for simplicity.
}
\label{contents_1}
\end{table}
In this table it is assumed that
 all twelve D6-branes of D6${}_4$
 are on top of one of eight O6-branes with the same winding numbers.
The same is also assumed
 for eight and twelve D6-branes of D6${}_5$ and D6${}_6$.

\begin{figure}
\begin{center}
\epsfig{file=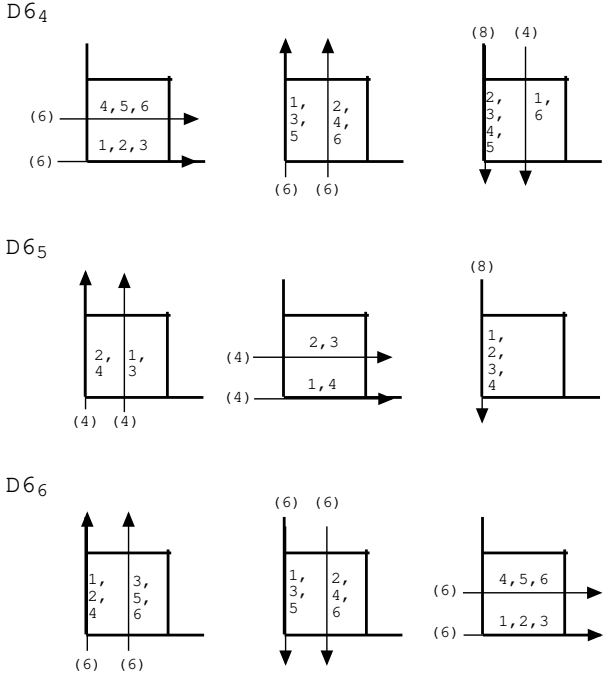,width=80mm}
\end{center}
\caption{
Configurations of twelve, eight and twelve D6-branes
 of D6${}_4$, D6${}_5$ and D6${}_6$ branes, respectively.
The numbers in brackets are multiplicities of D6-brane stacks,
 and the numbers without brackets specify
 one of the USp$(2)$ gauge groups in Eqs. (\ref{gauge_D6_4}),
 (\ref{gauge_D6_5}) and (\ref{gauge_D6_6}).
See footnote for details. 
}
\label{usp-conf}
\end{figure}

We break three
 USp$(12)_{{\rm D6}_4}$, USp$(8)_{{\rm D6}_5}$
 and USp$(12)_{{\rm D6}_6}$
 gauge symmetries
 to the factors of USp$(2)$ gauge symmetries
 by configuring D6-branes of D6${}_4$, D6${}_5$ and D6${}_6$
 as in Fig.\ref{usp-conf}.
The resultant gauge symmetries are respectively as follows.
\begin{equation}
 \mbox{USp}(12)_{{\rm D6}_4} \longrightarrow
 \directprod_{\alpha=1}^6 \mbox{USp}(2)_{{\rm D6}_4,\alpha},
\label{gauge_D6_4}
\end{equation}
\begin{equation}
 \mbox{USp}(8)_{{\rm D6}_5} \longrightarrow
 \directprod_{a=1}^4 \mbox{USp}(2)_{{\rm D6}_5,a},
\label{gauge_D6_5}
\end{equation}
\begin{equation}
 \mbox{USp}(12)_{{\rm D6}_6} \longrightarrow
 \directprod_{\alpha=1}^6 \mbox{USp}(2)_{{\rm D6}_6,\alpha}.
\label{gauge_D6_6}
\end{equation}
All of these USp$(2)$ gauge intersections
 can be naturally stronger than any other unitary gauge interactions.
If we choose $\kappa_4 M_s \sim 1$ and $\chi \sim 0.1$,
 where $\kappa_4=\sqrt{8 \pi G_N}$, $M_s = 1 / \sqrt{\alpha'}$,
 the dynamical scales of all USp$(2)$
 gauge interactions are of the order of $M_s$,
 and the values of the standard model gauge coupling constants
 are reasonably of ${\cal O}(0.01)$ at the string scale.
We call these strong USp$(2)$ interactions
 ``hypercolor'' interactions.
\footnote{
We have many vector-like matters in the configuration of Fig.2,
 since, for example, D6-branes of D6${}_4$ and D6${}_5$
 overlap in the third torus and intersect in other tori.
These vector-like matters
 spoil asymptotic freedom of ``hypercolor'' dynamics.
But, these vector-like matters get masses of Planck scale,
 if we move four D-branes of each USp$(2)$ (two D-branes and
 their images) a little away from orientifold planes in one torus
 in a consistent way with orientifold projections
 and avoid overlapping.
This procedure does not change the gauge symmetry and
 chiral matter contents of the model.
In the estimation of the Yukawa couplings,
 we do not consider this fine structure of the D-brane configuration,
 since it does not change our main results.
}

The field content
 in the intersecting D6${}_1$-D6${}_2$-D6${}_4$ sector
 (left-handed sector) is given in Table \ref{contents_left}.
\begin{table}
 \begin{tabular}{|c|c|c|}
  \hline
  sector             & ${}_{\left(
                         \mbox{SU}(3)_c \times \mbox{SU}(2)_L
                        \right)
                        \times
                        \left(
                         \mbox{USp}(2)_1 \times
                         \mbox{USp}(2)_2 \times
                        \right.}$
                     & field \\
                     & ${}_{\left.
                         \mbox{USp}(2)_3 \times
                         \mbox{USp}(2)_4 \times
                         \mbox{USp}(2)_5 \times                  
                         \mbox{USp}(2)_6                         
                        \right)_{D6_4}}$
                     & \\
                     & ($Y/2, Q_R$)($Q_L, Q_c+Q, Q_1+Q_2$)
                     & \\
  \hline\hline
  $D6_1 \cdot D6_2$  & $(3^*, 2)(1,1,1,1,1,1)_{(-1/6,0)(+1,-1,0)}
                         \times 2$
                     & ${\bar q}_i$ \\
                     & $(1, 2)(1,1,1,1,1,1)_{(+1/2,0)(+1,-1,0)}
                         \times 2$
                     & ${\bar l}_i$  \\
  \hline
  $D6_1 \cdot D6_4$  & $(1, 2)(2,1,1,1,1,1)_{(0,0)(-1,0,0)}$
                     & $D_\alpha$ \\
                     & $(1, 2)(1,2,1,1,1,1)_{(0,0)(-1,0,0)}$
                     & \\
                     & $(1, 2)(1,1,2,1,1,1)_{(0,0)(-1,0,0)}$
                     & \\
                     & $(1, 2)(1,1,1,2,1,1)_{(0,0)(-1,0,0)}$
                     & \\
                     & $(1, 2)(1,1,1,1,2,1)_{(0,0)(-1,0,0)}$
                     & \\
                     & $(1, 2)(1,1,1,1,1,2)_{(0,0)(-1,0,0)}$
                     & \\
  \hline
  $D6_2 \cdot D6_4$  & $(3, 1)(2,1,1,1,1,1)_{(+1/6,0)(0,+1,0)}$
                     & $C_\alpha$  \\
                     & $(3, 1)(1,2,1,1,1,1)_{(+1/6,0)(0,+1,0)}$
                     & \\
                     & $(3, 1)(1,1,2,1,1,1)_{(+1/6,0)(0,+1,0)}$
                     & \\
                     & $(3, 1)(1,1,1,2,1,1)_{(+1/6,0)(0,+1,0)}$
                     & \\
                     & $(3, 1)(1,1,1,1,2,1)_{(+1/6,0)(0,+1,0)}$
                     & \\
                     & $(3, 1)(1,1,1,1,1,2)_{(+1/6,0)(0,+1,0)}$
                     & \\
                     & $(1, 1)(2,1,1,1,1,1)_{(-1/2,0)(0,+1,0)}$
                     & $N_\alpha$  \\
                     & $(1, 1)(1,2,1,1,1,1)_{(-1/2,0)(0,+1,0)}$
                     & \\
                     & $(1, 1)(1,1,2,1,1,1)_{(-1/2,0)(0,+1,0)}$
                     & \\
                     & $(1, 1)(1,1,1,2,1,1)_{(-1/2,0)(0,+1,0)}$
                     & \\
                     & $(1, 1)(1,1,1,1,2,1)_{(-1/2,0)(0,+1,0)}$
                     & \\
                     & $(1, 1)(1,1,1,1,1,2)_{(-1/2,0)(0,+1,0)}$
                     & \\
  \hline
 \end{tabular}
\caption{
Field contents of the left-handed sector.
Here, $i=1,2$ and $\alpha = 1,2, \cdots ,6$.
}
\label{contents_left}
\end{table}
The confinement of
 six USp$(2)_{{\rm D6}_4,\alpha}$ gauge interactions
 gives six generations of left-handed quark and lepton doublets:
\begin{equation}
 C_\alpha D_\alpha \sim q_\alpha,
 \qquad
 N_\alpha D_\alpha \sim l_\alpha,
\end{equation}
 where $\alpha=1,2,\cdots,6$.
Two of these six left-handed quark doublets
 and two of these six left-handed lepton doublets
 become massive through the string-level Yukawa interactions of
\begin{equation}
 W_{\rm left} =
 \sum_{i,\alpha}
  g^{\rm left-q}_{i \alpha} {\bar q}_i C_\alpha D_\alpha
 +
 \sum_{i,\alpha}
  g^{\rm left-l}_{i \alpha} {\bar l}_i N_\alpha D_\alpha,
\label{yukawa_left}
\end{equation}
 where $i=1,2$.
We have four massless generations of
 left-handed quark and lepton doublets.
The values of masses are given as 
$g^{\rm left-q}_{i \alpha} \Lambda_L$ and 
$g^{\rm left-l}_{i \alpha} \Lambda_L$, where 
$\Lambda_L$ denotes 
 the dynamical scale of USp$(2)_{{\rm D6}_4,\alpha}$.
The detailed structure
 of the Yukawa coupling matrix $g^{\rm left-q}_{i \alpha}$
 is discussed in the next section.
Exactly the same happens in 
 the intersecting D6${}_2$-D6${}_3$-D6${}_6$ sector
 (right-handed sector).
The confinement of
 six USp$(2)_{{\rm D6}_6,\alpha}$ gauge interactions
 gives six generations of right-handed quarks and leptons:
\begin{eqnarray}
&
 {\bar C}_\alpha {\bar D}^{(-)}_\alpha \sim u_\alpha,
 \qquad
 {\bar C}_\alpha {\bar D}^{(+)}_\alpha \sim d_\alpha,
&
\nonumber\\
&
 {\bar N}_\alpha {\bar D}^{(-)}_\alpha \sim \nu_\alpha,
 \qquad
 {\bar N}_\alpha {\bar D}^{(+)}_\alpha \sim e_\alpha.
&
\end{eqnarray}
We have four massless generations of
 right-handed quarks and leptons
 through the string-level Yukawa interactions of
\begin{eqnarray}
 W_{\rm right}
 & = &
 \sum_{i,\alpha}
 g^{\rm right-u}_{i \alpha} {\bar u}_i {\bar C} {\bar D}^{(-)}
 +
 \sum_{i,\alpha}
 g^{\rm right-d}_{i \alpha} {\bar d}_i {\bar C} {\bar D}^{(+)}
\nonumber\\
 & + &
 \sum_{i,\alpha}
 g^{{\rm right-}\nu}_{i \alpha} {\bar \nu}_i {\bar N} {\bar D}^{(-)}
 +
 \sum_{i,\alpha}
 g^{\rm right-e}_{i \alpha} {\bar e}_i {\bar N} {\bar D}^{(+)}.
\label{yukawa_right}
\end{eqnarray}
The detailed structure
 of the Yukawa coupling matrices
 $g^{\rm right-u}_{i \alpha}$ and $g^{\rm right-d}_{i \alpha}$
 is also discussed in the next section.
The values of masses are given as 
$g^{\rm right-u}_{i \alpha} \Lambda_R$ and 
$g^{\rm right-d}_{i \alpha} \Lambda_R$ with 
the dynamical scale of  USp$(2)_{{\rm D6}_6,\alpha}$.
Here, we stress that
 the origin of generation in this model
 is not the multiple intersections of D6-branes,
 but the number of different D6-branes with the same winding numbers.

The similar happens in 
 the intersecting D6${}_1$-D6${}_3$-D6${}_5$ sector
 (Higgs sector).
The confinement of
 two USp$(2)_{{\rm D6}_5,a}$ gauge interactions
 gives eight composite Higgs fields:
\begin{equation}
 T_a T^{(+)} \sim H^{(2)}_a,
 \qquad
 T_a T^{(-)} \sim {\bar H}^{(1)}_a,
 \end{equation}
 where $a=1,2,3,4$.
Four of these eight composite Higgs fields become massive
 through the string-level Yukawa interactions of
\begin{equation}
 W_{\rm higgs} =
 \sum_{i,a} g^{(1)}_{ia} H_i^{(1)} T_a T^{(-)}_a
 +
 \sum_{i,a} g^{(2)}_{ia} {\bar H}_i^{(2)} T_a T^{(+)}_a.
\label{yukawa_higgs}
\end{equation}
The masses of four pairs of Higgs fields are given as
 $g^{(1)}_{ia} \Lambda_H$ and $g^{(2)}_{ia} \Lambda_H$,
 where $\Lambda_H$ is the scale of dynamics
 of USp$(2)_{{\rm D6}_5,a}$.
Two pairs of composite Higgs fields remain massless.

 From Fig.\ref{intersec}
 we see the existence of the following interactions.
\begin{eqnarray}
 W_{\rm 6-fields}
 & = &
 \sum_{\alpha,\beta=1}^6 \sum_{a=1}^4
 {{g^u_{\alpha \beta a}} \over {M_s^3}}
  [ C_\alpha D_\alpha ] 
  [ {\bar C}_\beta {\bar D}^{(-)}_\beta ]
  [ T_a T^{(+)}_a ]
\nonumber\\
 & + &
 \sum_{\alpha,\beta=1}^6 \sum_{a=1}^4
 {{g^d_{\alpha \beta a}} \over {M_s^3}}
  [ C_\alpha D_\alpha ]
  [ {\bar C}_\beta {\bar D}^{(+)}_\beta ]
  [ T_a T^{(-)}_a ]
\nonumber\\
 & + &
 \sum_{\alpha,\beta=1}^6 \sum_{a=1}^4
 {{g^\nu_{\alpha \beta a}} \over {M_s^3}}
  [ N_\alpha D_\alpha ]
  [ {\bar N}_\beta {\bar D}^{(-)}_\beta ]
  [ T_a T^{(+)}_a ]
\nonumber\\
 & + &
 \sum_{\alpha,\beta=1}^6 \sum_{a=1}^4
 {{g^e_{\alpha \beta a}} \over {M_s^3}}
  [ N_\alpha D_\alpha ]
  [ {\bar N}_\beta {\bar D}^{(+)}_\beta ]
  [ T_a T^{(-)}_a ].
\label{6-fields}
\end{eqnarray}
After the ``hypercolor'' confinement
 these interactions give Yukawa interactions
 for the mass generation of quarks and leptons.
The Yukawa coupling matrices are given by
\begin{eqnarray}
 y^u_{\alpha \beta a} & \simeq &
  g^u_{\alpha \beta a}
  {{\Lambda_L \Lambda_R \Lambda_H} \over {M_s^3}}
  \sim g^u_{\alpha \beta a},
\label{yukawa-u}
\\
 y^d_{\alpha \beta a} & \simeq &
  g^d_{\alpha \beta a}
  {{\Lambda_L \Lambda_R \Lambda_H} \over {M_s^3}}
  \sim g^d_{\alpha \beta a},
\label{yukawa-d}
\\
 y^\nu_{\alpha \beta a} & \simeq &
  g^\nu_{\alpha \beta a}
  {{\Lambda_L \Lambda_R \Lambda_H} \over {M_s^3}}
  \sim g^\nu_{\alpha \beta a},
\label{yukawa-nu}
\\
 y^e_{\alpha \beta a} & \simeq &
  g^e_{\alpha \beta a}
  {{\Lambda_L \Lambda_R \Lambda_H} \over {M_s^3}}
  \sim g^e_{\alpha \beta a},
\label{yukawa-e}
\end{eqnarray}
 since all the scales of dynamics,
 $\Lambda_L$, $\Lambda_R$ and $\Lambda_H$
 are of the order of the string scale $M_s$.
The detailed structure of these Yukawa coupling matrices
 $g^u_{\alpha \beta a}$ and $g^d_{\alpha \beta a}$
 is investigated in the next section.

\section{The structure of Yukawa Coupling Matrices}
\label{sec:structure}

Here 
 we investigate the structure of Yukawa coupling matrices
 for quark masses of the model introduced in the previous section.
We do not require
 the realistic $\mu$-term (realistic value of the Higgs mass),
 since the aim of the analysis is simply to show the possibility
 of having nontrivial structure of Yukawa coupling matrices, 
e.g. non-vanishing mixing angles.
The possibility of
 entire mass generation of quarks and leptons in this kind of models
 will be investigated in future works.

First,
 we evaluate the Yukawa coupling matrices
 $g^{\rm left-q}_{i \alpha}$ in Eq.(\ref{yukawa_left}) and
 $g^{\rm right-u}_{i \alpha}$ and $g^{\rm right-d}_{i \alpha}$
 in Eq.(\ref{yukawa_right}) by using Eq.(\ref{Yukawa-cl}).
Three Yukawa-interacting fields
 are localized at three intersection points of these three D6-branes.
The Yukawa couplings, $g^{\rm left-q}_{i \alpha}$, 
 $g^{\rm right-u}_{i \alpha}$ and $g^{\rm right-d}_{i \alpha}$, 
are  obtained as 
\begin{eqnarray}
 g^{\rm left-q} & = &
  \left(
  \begin{array}{cccccc}
  \varepsilon_3 &
  \varepsilon_2 &
  1 &
  \varepsilon_1^2 \varepsilon_2 &
  \varepsilon_1^2 &
  \varepsilon_1^2 \varepsilon_2 \varepsilon_3
  \\
  \varepsilon_1^2 \varepsilon_3 &
  \varepsilon_1^2 \varepsilon_2 & 
  \varepsilon_1^2 &
  \varepsilon_2 &
  1 &
  \varepsilon_2 \varepsilon_3
  \end{array}
  \right),
\label{mass-q}\\
 g^{\rm right-u} & = &
  \left(
  \begin{array}{cccccc}
  1 &
  \varepsilon_2 &
  \varepsilon_1 &
  \varepsilon_2 \varepsilon_3^2 & 
  \varepsilon_1 \varepsilon_3^2 &
  \varepsilon_1 \varepsilon_2 \varepsilon_3^2
  \\
  \varepsilon_3^2 &
  \varepsilon_2 \varepsilon_3^2 & 
  \varepsilon_1 \varepsilon_3^2 & 
  \varepsilon_2 & \varepsilon_1 &
  \varepsilon_1 \varepsilon_2
  \end{array}
  \right),
\label{mass-u}\\
 g^{\rm right-d} & = &
  \left(
  \begin{array}{cccccc}
  \varepsilon_1 & \varepsilon_1 \varepsilon_2 &
  1 & 
  \varepsilon_1 \varepsilon_2 \varepsilon_3^2 &
  \varepsilon_3^2 &
  \varepsilon_2 \varepsilon_3^2
  \\
  \varepsilon_1 \varepsilon_3^2 &
  \varepsilon_1 \varepsilon_2 \varepsilon_3^2 & 
  \varepsilon_3^2 & 
  \varepsilon_1 \varepsilon_2 &
  1 &
  \varepsilon_2
  \end{array}
  \right),   
\label{mass-d}
\end{eqnarray}
 where $\varepsilon_i = \exp (- A_i / 2 \pi \alpha')$ and 
 $A_i$ is the $1/8$ of the area of the $i$-th torus. 
In general, classical solutions with larger area also contribute 
to the Yukawa couplings.
However, we have used the approximation that the classical 
action corresponding to the minimum area contributes 
dominantly to the Yukawa couplings.
In the following
 we assume that all $A_i$ are larger than $2 \pi \alpha'$,
 namely $\varepsilon < 1$.
 From these results
 we can say in good approximation
 that the massless left-handed quark doublets
 are $q_1$, $q_2$, $q_4$ and $q_6$,
 and the massless right-handed down-type quarks
 are $d_1$, $d_2$, $d_4$ and $d_6$.
For right-handed up-type quarks,
 $u_2$, $u_3$, $u_6$ and one linear combination of $u_4$ and $u_5$
 are approximately massless.
Especially,
 in case of $\varepsilon_1 \ll \varepsilon_2$,
 $u_2$, $u_3$, $u_5$ and $u_6$ are approximately massless,
 and in case of $\varepsilon_2 \ll \varepsilon_1$,
 $u_2$, $u_3$, $u_4$ and $u_6$ are approximately massless.

Although it is not easy to calculate the 6-point coupling matrices of
 $g^u_{\alpha \beta a}$ and $g^d_{\alpha \beta a}$
 in Eq.(\ref{6-fields}), 
generic n-point functions have been analyzed 
in Ref.~\cite{Abel:2003yx}.
Our purpose is not to realize precisely realistic fermion masses 
and mixing angles, but 
to show that a new type of nontrivial Yukawa matrices can 
be derived in our scenario, e.g. suppressed but non-vanishing mixing angles.
For such purpose, an approximation, which is 
reliable to order estimation on suppression factors, is sufficient.
The six-point function consists of classical and quantum 
contributions, i.e. $Z_{cl} Z_{q}$, like the 3-point and 
4-point functions.
The classical part $Z_{cl}$ is important for our purpose because 
that can lead to a suppression factor, while 
the quantum part is expected to be of O(1) and 
does not contribute to lead to a suppression factor.
The classical part is written as 
$Z_{cl}=\sum_{X_{cl}}e^{-S_{cl}}$ like Eq.(\ref{Yukawa-cl}), 
where $X_{cl}$ the classical string solutions, 
which have the asymptotic behavior corresponding to 
local open string modes near intersecting points.
The minimum action corresponds to the sum of the areas of 
hexagons with six intersecting points 
on each torus \cite{Abel:2003yx}.
Hence, in order to give an order estimation of 
 6-point couplings 
$g^{u,d}_{\alpha \beta a}$, 
 we use the approximation that $g^{u,d}_{\alpha \beta a} 
\sim \prod_i e^{-\Sigma_i/2\pi \alpha'}$, 
which is the same as Eq.(\ref{Yukawa-cl}) except 
 simply replacing the triangular areas $\Sigma_i$
 by the hexagonal areas.
This approximation gives a sufficient order estimation of 
the exponential suppression enough for our purpose,
 though some sub-leading corrections are neglected.
Actually, in our model, all the hexagons reduce to triangles,
because some intersecting points always coincide and only
three of the six sit at the different places.
Hence, we obtain the following result.
\begin{eqnarray}
 & g^u_{\alpha \beta a=1} & = g^d_{\alpha \beta a=1} 
\nonumber\\
 & = &
\left(
\begin{array}{cccccc}
\varepsilon_1 \varepsilon_3 &
\varepsilon_1 \varepsilon_2 \varepsilon_3 & 
\varepsilon_1^2 \varepsilon_3 &
\varepsilon_1 \varepsilon_2 \varepsilon_3 & 
\varepsilon_1^2 \varepsilon_3 &
\varepsilon_1^2 \varepsilon_2 \varepsilon_3
\cr
\varepsilon_1 \varepsilon_2 &
\varepsilon_1 \varepsilon_2^2 &
\varepsilon_1^2 \varepsilon_2 &
\varepsilon_1 \varepsilon_2^2 \varepsilon_3^2 & 
\varepsilon_1^2 \varepsilon_2 \varepsilon_3^2 &
\varepsilon_1^2 \varepsilon_2^2 \varepsilon_3^2
\cr
\varepsilon_1 &
\varepsilon_1 \varepsilon_2 &
\varepsilon_1^2 & 
\varepsilon_1 \varepsilon_2 \varepsilon_3^2 & 
\varepsilon_1^2 \varepsilon_3^2 &
\varepsilon_1^2 \varepsilon_2 \varepsilon_3^2
\cr
\varepsilon_1 \varepsilon_2 &
\varepsilon_1 \varepsilon_2^2 &
\varepsilon_2 &
\varepsilon_1 \varepsilon_2^2 \varepsilon_3^2 & 
\varepsilon_2 \varepsilon_3^2 &
\varepsilon_2^2 \varepsilon_3^2
\cr
\varepsilon_1 &
\varepsilon_1 \varepsilon_2 &
1 & 
\varepsilon_1 \varepsilon_2 \varepsilon_3^2 & 
\varepsilon_3^2 &
\varepsilon_2 \varepsilon_3^2
\cr
\varepsilon_1 \varepsilon_2 \varepsilon_3 & 
\varepsilon_1 \varepsilon_2^2 \varepsilon_3 & 
\varepsilon_2 \varepsilon_3 & 
\varepsilon_1 \varepsilon_2^2 \varepsilon_3 &
\varepsilon_2 \varepsilon_3 &
\varepsilon_2^2 \varepsilon_3
\end{array}
\right)_{\alpha\beta},
 \\
 & g^u_{\alpha \beta a=2} & = g^d_{\alpha \beta a=2}
\nonumber\\
 & = &
\left(
\begin{array}{cccccc}
\varepsilon_2^2 \varepsilon_3 &
\varepsilon_2 \varepsilon_3 & 
\varepsilon_1 \varepsilon_2^2 \varepsilon_3 &
\varepsilon_2 \varepsilon_3 & 
\varepsilon_1 \varepsilon_2^2 \varepsilon_3 &
\varepsilon_1 \varepsilon_2 \varepsilon_3
\cr
\varepsilon_2 &
1 &
\varepsilon_1 \varepsilon_2 &
\varepsilon_3^2 & 
\varepsilon_1 \varepsilon_2 \varepsilon_3^2 &
\varepsilon_1 \varepsilon_3^2
\cr
\varepsilon_2^2 &
\varepsilon_2 &
\varepsilon_1 \varepsilon_2^2 & 
\varepsilon_2 \varepsilon_3^2 & 
\varepsilon_1 \varepsilon_2^2 \varepsilon_3^2 &
\varepsilon_1 \varepsilon_2 \varepsilon_3^2
\cr
\varepsilon_1^2 \varepsilon_2 &
\varepsilon_1^2 &
\varepsilon_1 \varepsilon_2 &
\varepsilon_1^2 \varepsilon_3^2 & 
\varepsilon_1 \varepsilon_2 \varepsilon_3^2 &
\varepsilon_1 \varepsilon_3^2
\cr
\varepsilon_1^2 \varepsilon_2^2 &
\varepsilon_1^2 \varepsilon_2 &
\varepsilon_1 \varepsilon_2^2 & 
\varepsilon_1^2 \varepsilon_2 \varepsilon_3^2 & 
\varepsilon_1 \varepsilon_2^2 \varepsilon_3^2 &
\varepsilon_1 \varepsilon_2 \varepsilon_3^2
\cr
\varepsilon_1^2 \varepsilon_2 \varepsilon_3 & 
\varepsilon_1^2 \varepsilon_3 & 
\varepsilon_1 \varepsilon_2 \varepsilon_3 & 
\varepsilon_1^2 \varepsilon_3 &
\varepsilon_1 \varepsilon_2 \varepsilon_3 &
\varepsilon_1 \varepsilon_3
\end{array}
\right)_{\alpha\beta},
 \\
 & g^u_{\alpha \beta a=3} & = g^d_{\alpha \beta a=3}
\nonumber\\
 & = &
\left(
\begin{array}{cccccc}
\varepsilon_1 \varepsilon_2^2 \varepsilon_3 &
\varepsilon_1 \varepsilon_2 \varepsilon_3 & 
\varepsilon_1^2 \varepsilon_2^2 \varepsilon_3 &
\varepsilon_1 \varepsilon_2 \varepsilon_3 & 
\varepsilon_1^2 \varepsilon_2^2 \varepsilon_3 &
\varepsilon_1^2 \varepsilon_2 \varepsilon_3
\cr
\varepsilon_1 \varepsilon_2 &
\varepsilon_1 &
\varepsilon_1^2 \varepsilon_2 &
\varepsilon_1 \varepsilon_3^2 & 
\varepsilon_1^2 \varepsilon_2 \varepsilon_3^2 &
\varepsilon_1^2 \varepsilon_3^2
\cr
\varepsilon_1 \varepsilon_2^2 &
\varepsilon_1 \varepsilon_2 &
\varepsilon_1^2 \varepsilon_2^2 & 
\varepsilon_1 \varepsilon_2 \varepsilon_3^2 & 
\varepsilon_1^2 \varepsilon_2^2 \varepsilon_3^2 &
\varepsilon_1^2 \varepsilon_2 \varepsilon_3^2
\cr
\varepsilon_1 \varepsilon_2 &
\varepsilon_1 &
\varepsilon_2 &
\varepsilon_1 \varepsilon_3^2 & 
\varepsilon_2 \varepsilon_3^2 &
\varepsilon_3^2
\cr
\varepsilon_1 \varepsilon_2^2 &
\varepsilon_1 \varepsilon_2 &
\varepsilon_2^2 & 
\varepsilon_1 \varepsilon_2 \varepsilon_3^2 & 
\varepsilon_2^2 \varepsilon_3^2 &
\varepsilon_2 \varepsilon_3^2
\cr
\varepsilon_1 \varepsilon_2 \varepsilon_3 & 
\varepsilon_1 \varepsilon_3 & 
\varepsilon_2 \varepsilon_3 & 
\varepsilon_1 \varepsilon_3 &
\varepsilon_2 \varepsilon_3 &
\varepsilon_3
\end{array}
\right)_{\alpha\beta},
 \\
 & g^u_{\alpha \beta a=4} & = g^d_{\alpha \beta a=4}
\nonumber\\
 & = &
\left(
\begin{array}{cccccc}
\varepsilon_3 &
\varepsilon_2 \varepsilon_3 & 
\varepsilon_1 \varepsilon_3 &
\varepsilon_2 \varepsilon_3 & 
\varepsilon_1 \varepsilon_3 &
\varepsilon_1 \varepsilon_2 \varepsilon_3
\cr
\varepsilon_2 &
\varepsilon_2^2 &
\varepsilon_1 \varepsilon_2 &
\varepsilon_2^2 \varepsilon_3^2 & 
\varepsilon_1 \varepsilon_2 \varepsilon_3^2 &
\varepsilon_1 \varepsilon_2^2 \varepsilon_3^2
\cr
1 &
\varepsilon_2 &
\varepsilon_1 & 
\varepsilon_2 \varepsilon_3^2 & 
\varepsilon_1 \varepsilon_3^2 &
\varepsilon_1 \varepsilon_2 \varepsilon_3^2
\cr
\varepsilon_1^2 \varepsilon_2 &
\varepsilon_1^2 \varepsilon_2^2 &
\varepsilon_1 \varepsilon_2 &
\varepsilon_1^2 \varepsilon_2^2 \varepsilon_3^2 & 
\varepsilon_1 \varepsilon_2 \varepsilon_3^2 &
\varepsilon_1 \varepsilon_2^2 \varepsilon_3^2
\cr
\varepsilon_1^2 &
\varepsilon_1^2 \varepsilon_2 &
\varepsilon_1 & 
\varepsilon_1^2 \varepsilon_2 \varepsilon_3^2 & 
\varepsilon_1 \varepsilon_3^2 &
\varepsilon_1 \varepsilon_2 \varepsilon_3^2
\cr
\varepsilon_1^2 \varepsilon_2 \varepsilon_3 & 
\varepsilon_1^2 \varepsilon_2^2 \varepsilon_3 & 
\varepsilon_1 \varepsilon_2 \varepsilon_3 & 
\varepsilon_1^2 \varepsilon_2^2 \varepsilon_3 &
\varepsilon_1 \varepsilon_2 \varepsilon_3 &
\varepsilon_1 \varepsilon_2^2 \varepsilon_3
\end{array}
\right)_{\alpha\beta}.
\end{eqnarray}
Since we have six pair of Higgs fields
 and only the composite Higgs fields have Yukawa couplings,
 there are eight Yukawa coupling matrices for quark masses.

All of these Yukawa matrices are nontrivial in the sense that 
they differ from the factorizable form of Eq.(\ref{yukawa-fact}).
Detail calculation is necessary for full analysis on 
these Yukawa matrices by varying parameters.
However, our purpose is not to analyze this model in detail, 
but to show the possibility for leading to nontrivial results 
in our scenario, i.e. non-vanishing mass ratios and mixing angles, 
which can be realistic.
Therefore, here we show an simple case.

Now, let us consider the case with $\varepsilon_2 \ll \varepsilon_1$.
In this case, the modes $u_1$ and $u_5$ as well as 
$q_3$, $q_5$, $d_3$ and $d_5$ dominantly gain large masses 
through the interactions of
 Eqs.(\ref{yukawa_left}) and (\ref{yukawa_right})
 with coupling matrices of
 Eqs.(\ref{mass-q}), (\ref{mass-u}) and (\ref{mass-d}).
The other modes approximately correspond to four generations.
For example, let us consider the Yukawa matrices,
$g^u_{a=2}$ and $g^{d}_{a=3}$, that is, we assume that 
one pair of composite Higgs doublet fields,
 $({\bar H}^{(1)}_3, H^{(2)}_2)$, is dominant to generate quark mass 
matrices at the weak scale.
Then we consider the parameter region that 
$\varepsilon_2 \ll \varepsilon_3 \ll \varepsilon_1 (\sim 1)$, 
and in this parameter region the matter fields $(q_2, q_4, q_6)$, 
 $(u_2,u_6)$ and $(d_2,d_6)$ have large entries 
in the Yukawa matrices $g^u_{a=2}$ and $g^{d}_{a=3}$.
Here, let us concentrate ourselves to two heavy modes among four generations.
The $(2 \times 2)$ submatrices with large entries 
in the Yukawa matrices $g^u_{a=2}$ and $g^{d}_{a=3}$ 
are written  up to $O(\varepsilon_3^2)$ as 
\begin{eqnarray}
g^u_{a=2} & \rightarrow  &
 \left(
  \begin{array}{cc}
  1 &
  0
  \cr
  \varepsilon_1^2 \varepsilon_3 & 
  \varepsilon_1 \varepsilon_3
  \end{array}
 \right),
 \label{u-2-22}
\end{eqnarray}
for the up sector, 
\begin{eqnarray}
  g^d_{a=3} & \rightarrow &
 \left(
  \begin{array}{cccc}
  \varepsilon_1 &
  0
  \cr
  \varepsilon_1 \varepsilon_3 & 
  \varepsilon_3
  \end{array}
 \right),
 \label{d-3-22}
 \end{eqnarray}
for the down sector in the field basis 
$((q_2 + q_4)/\sqrt{2},q_6)$, $(u_2,u_6)$, $(d_2,d_6)$.

Now we can calculate mass eigenvalues of two heavy modes 
among four generations, i.e. $m_{u,3}$ and $m_{u,4}$ 
for the up sector and $m_{d,3}$ and $m_{d,4}$ 
for the down sector, and their mixing angle $V_{34}$.
The mass ratios and the mixing angle is obtained as 
\begin{equation}
\frac{m_{u,3}}{m_{u,4}} \approx \varepsilon_1 \varepsilon_3, \qquad 
\frac{m_{d,3}}{m_{d,4}} \approx \frac{\varepsilon_3}{ \varepsilon_1}, \qquad 
V_{34} \approx \varepsilon_3 ,
\end{equation}
that is, we have the following relation,
\begin{equation}
\sqrt{\frac{m_{u,3}}{m_{u,4}} \frac{m_{d,3}}{m_{d,4}}}
\approx 
V_{34} ,
\label{mass-r-V}
\end{equation}
at the composite scale.

It is interesting to compare these results
 with the experimental values of quark mass ratios,
 $\frac{m_c}{m_t}$ and $\frac{m_s}{m_b}$,
 and the mixing angle $V_{cb}$.
At the weak scale, the experimental values of mass ratios, 
$\frac{m_c}{m_t} = 0.0038$ and $\frac{m_s}{m_b} = 0.025$, 
lead to 
\begin{equation}
\sqrt{\frac{m_c}{m_t}\frac{m_s}{m_b}} = 0.01,
\end{equation}
and the mixing angle is 
\begin{equation}
V_{cb} = 0.04 .
\end{equation}
We find that the values of parameters
 $\varepsilon_1 \sim 0.5$ and $\varepsilon_3 \sim 0.01$
 lead to almost realistic structure of 
 quark Yukawa coupling matrices.
Note that our approximation is reliable as order estimation.

We have shown that 
 the structure of the Yukawa coupling matrices 
 at least for the $(2 \times 2)$ sub-matrices 
 can be realistic in the simple case through 
 our scenario of the dynamical generation
 of Yukawa coupling matrices.
The most relevant fact is the different origin of the generation
 from that in the conventional intersecting D-brane models.
The origin of the generation
 is not the multiple intersection of D-branes,
 but many different D-branes with the same multiplicity and
 the same winding numbers.

We give several comments in order before closing this section.

In this model
 a large Yukawa coupling is obtained in case of which
 all six localization points of ``preons'' are coincides
 in all three tori.
Therefore, 
 it seems very difficult to obtain the Yukawa coupling
 of the order of unity.
But, we have shown that
 it is possible and can accidentally happen
 (see Eq. (\ref{u-2-22})).

For more rigorous investigations,
 we should take care of the normalization of fields.
The low-energy effective fields in string theory
 should be normalized by considering the moduli dependence
 of the K\"ahler potential.
A concrete analysis
 on the moduli dependence of the K\"ahler potential
 in type IIA intersecting D-brane models
 is given in Ref.\cite{LMRS}.
The normalization of the low-energy effective field
 after ``hypercolor'' confinement should also be considered
 more precisely.
Unfortunately,
 there is no established method
 to obtain the K\"ahler potentials for composite fields
 in strong coupling gauge theories in four dimensions,
 except for some special cases\cite{ADS,Poppitz-Randall}.

\section{Conclusions}
\label{sec:conclusions}

The structure of Yukawa coupling matrices
 in models from type IIA $T^6/({\rm Z}_2 \times {\rm Z}_2)$
 orientifolds ($T^6 = T^2 \times T^2 \times T^2$)
 with intersecting D-branes has been investigated.
There is a difficulty
 to have a realistic Yukawa coupling matrices in the models
 in which the generation structure of quarks and leptons
 is originated from the multiple intersection of D-branes.
On the other hand,
 it has been shown that the structure of Yukawa coupling matrices
 in the models with dynamical generation of Yukawa coupling matrices
 is nontrivial.
Indeed, realistic values of the mixing angles $V_{cb}$ and 
mass ratios $m_c/m_t$ and $m_s/m_b$ can be realized.
The most relevant fact is the different origin of the generation.
The origin of the generation
 is not the multiple intersection of D-branes,
 but many different D-branes with the same multiplicity and
 the same winding numbers.

Here we give a comment on the origin of CP violating phases.
Inclusion of Wilson lines is one source of CP phases in 
Yukawa matrices \cite{CIM}.
As another source, the holomorphic dynamical scales
 $\Lambda_{R,L,H}$ may provide CP phases
 in effective Yukawa couplings in this class of models.


Although the possibility
 of having realistic Yukawa coupling matrices,
 this class of models with dynamical generation
 of Yukawa coupling matrices
 have many other phenomenological issues to be solved:
 constructing models with three generations and
 light Higgs pairs with realistic $\mu$-term
 for electroweak symmetry breaking,
 realization of the supersymmetry breaking,
 and so on.
In addition to these which are common
 in all the supersymmetric type IIA intersecting D-brane models,
 composite models usually have a problem of rapid proton decay.
It would be very interesting to explorer more realistic models
 by incorporating the results of efforts of the model buildings
 based on the field theory
 into the framework of the intersecting D-brane model.

\begin{acknowledgement}
The authors would like to thank Tatsuya Noguchi 
for useful discussions.
N.~K., T.~K. and N.~O. would like to
thank hospitality of RIKEN, where
a part of this work was studied.
T.~K.\/ is supported in part by the Grant-in-Aid for
Scientific Research (\#16028211) and
the Grant-in-Aid for
the 21st Century COE ``The Center for Diversity and
Universality in Physics'' from the Ministry of Education, Culture,
Sports, Science and Technology of Japan.
N.~M.\/ is supported by the Special Postdoctoral Researchers
Program at RIKEN.
N.~O.\/ is supported in part by the Grant-in-Aid for
Scientific Research (\#15740164).
\end{acknowledgement}

\end{document}